\documentclass[12 pt,aps,prd,showpacs]{revtex4-1}   
\usepackage{natbib}
\usepackage{amsmath}    
\usepackage{graphicx}   
\usepackage{epsfig}
\def\b{\begin{equation}}
\def\e{\end{equation}}

\begin{document}
\title{Relativistic bands in the spectrum of created particles via dynamical Casimir effect}

\author{Andreson L. C. Rego$^{1}$, Jo\~ao Paulo da S. Alves$^{2}$, Danilo T. Alves$^{2}$ and C. Farina$^{1}$}
\affiliation{
$1$ - Instituto de F\'\i sica, Universidade Federal do Rio de Janeiro,
21945-970,
Rio de Janeiro, RJ, Brazil
\\
$2$ - Faculdade de F\'\i sica, Universidade Federal do
Par\'a,
66075-110,
Bel\'em, PA,  Brazil
}
\date{\today}
\begin{abstract}
We present a general analytical approach to investigate relativistic corrections in the dynamical Casimir effect (DCE). Particularly, we discuss the behavior of the additional frequency bands that appear in the
spectral distribution of the created particles when the first relativistic corrections are taken into account.
We do that in the context of circuit QED, by analyzing the setup used in the first measurement of the DCE, where a system with a time-dependent boundary condition simulates the mechanical motion of a single mirror in 1+1 dimensions. Our method is applicable to a large class of systems with oscillatory time-dependent parameters and can be generalized to higher dimensions and other fields.
\end{abstract}
%
\maketitle

\section{Introduction}

The theoretical prediction of the dynamical Casimir effect (DCE), essentially the conversion of vacuum fluctuations into real field excitations caused by moving mirrors, was made by Moore in 1970 \cite{Moore-1970}. This phenomenon was also investigated in other pioneering works by DeWitt \cite{Dewitt-PhysRep-1975} and  Fulling and Davies \cite{Fulling-Davies-PRS-1976-I, Fulling-Davies-PRS-1977-I}.
Soon it was realized that, in real experimental situations, the creation of photons via DCE
using mechanical motions of material plates is negligible, since the highest velocity that a mirror can achieve under 
la\-bo\-ra\-to\-ry conditions is very small in
comparison with the speed of light \cite{Moore-1970,Dodonov-PRA-1996,Lambrecht-et-al-EPJD-1998}.
A way to circumvent this difficulty is to simulate a moving mirror by a physical mechanism which gives rise to a time-dependent 
boundary condition (BC) for the field at a fixed mirror, an ingenious idea, first proposed by Yablonovitch in 1989 \cite{Yablonovitch-1989}. Several ex\-pe\-ri\-men\-tal proposals for the detection of the DCE are based on the simulation of moving boundaries \cite{Braggio-Agnesi,J-R-Johansson-G-Johansson-C-Wilson-F-Nori-PRL-2009-2010,
Dezael-Lambrecht-EPL-2010,Kawakubo-Yamamoto-PRA-2011,Faccio-Carusotto-EPL-2011}. 
We should also mention an ingenious proposal involving real mechanical motion of boundaries suggested by Kim {\it et al} 
\cite{Kim-Brownell-Onofrio-2006}, in which the dynamical Casimir photons would trigger a superradiance process which would then be observed.
One of these experimental proposals, based on a superconducting coplanar waveguide terminated by a SQUID (Superconducting Quantum Interference Device),
led to the announcement by Wilson {\it et al} of the first observation of the DCE \cite{Wilson-Nature-2011}.
In this experiment, a time-dependent magnetic flux is applied to the SQUID, changing 
its effective inductance, 
resulting in a time-dependent BC,
such that the coplanar waveguide becomes equi\-va\-lent to a one-dimensional transmission line with
variable length. This setup simulates a single moving mirror whose effective velocity
can achieve approximately $10\%$ of the speed of light \cite{J-R-Johansson-G-Johansson-C-Wilson-F-Nori-PRL-2009-2010}.

The DCE for a one-dimensional model with a single mirror moving in vacuum, with oscillatory motion
with frequency $\omega_0$, small amplitude and non-relativistic ve\-lo\-ci\-ties, was investigated
by Lambrecht {\it et al} \cite{Lambrecht-Jaekel-Reynaud-PRL-1996}.
These authors concluded that the spectral distribution of the created photons has a parabolic shape, with a maximum
at $\omega_0/2$ and no particle created with frequencies higher than $\omega_0$ \cite{Lambrecht-Jaekel-Reynaud-PRL-1996}.
In contrast, if the oscillatory motion has relativistic velocities,
the same authors pointed out the presence of additional frequency bands
in the spectrum of the crea\-ted particles. These bands vanish for all frequencies $\omega$
equal to an integer multiple of $\omega_0$, so that the spectrum is decomposed into a succession
of arches, each one limited by two successive multiples of $\omega_0$ \cite{Lambrecht-et-al-EPJD-1998}.
 At that time, this model was considered ``not realistic as it would imply a mirror's mechanical velocity appreciable compared to the speed of light'' \cite{Lambrecht-et-al-EPJD-1998}. 
  
However, in the recent experiment made by Wilson {\it et al} \cite{Wilson-Nature-2011}, the effective velocities can achieve $\sim 10\%$ of the velocity of light. This fact motivated us to investigate relativistic effects in the particle creation via DCE, particularly, the emergence of an additional and experimentally detectable band in the spectral distribution. This additional band provides 
an extra signature for identifying dynamical Casimir photons.
In the SQUID experiment, the DCE is modeled by a Robin BC at a fixed point but with a time-dependent Robin parameter. For this case, the relativistic
corrections (additional bands) were theo\-re\-ti\-cal\-ly predicted by Johansson {\it et al}
\cite{J-R-Johansson-G-Johansson-C-Wilson-F-Nori-PRL-2009-2010}.
In contrast with the model investigated in Ref. \cite{Lambrecht-et-al-EPJD-1998}, the spectrum
found in Ref. \cite{J-R-Johansson-G-Johansson-C-Wilson-F-Nori-PRL-2009-2010} does not
vanish for frequencies $\omega$ equal to multiples of $\omega_0$, and each higher order
band has a parabolic form going from 0 to an integer multiple of $\omega_0$. Robin BC in the DCE with moving plates were firstly investigated by Mintz {\it et al} \cite{Mintz-Farina-Maia-Neto-Robson-JPA-2006}. A generalization to 3+1 dimensions was recently reported in the literature \cite{{Rego-Mintz-Farina-Alves-PRD-2013}}. For Robin BC at fixed plates with time-dependent Robin parameter see Silva-Farina \cite{Silva-Farina-PRD-2011} and Fosco {\it et al} \cite{Fosco-et-al-PRD-2013}.

The shape of the additional bands investigated in Refs.
\cite{Lambrecht-et-al-EPJD-1998} and
\cite{J-R-Johansson-G-Johansson-C-Wilson-F-Nori-PRL-2009-2010} are different, and
also were calculated via different approaches, namely: an analytical approach
valid for a certain class of motions in $1+1$ dimensions \cite{Lambrecht-et-al-EPJD-1998} and
a computer numerical method \cite{J-R-Johansson-G-Johansson-C-Wilson-F-Nori-PRL-2009-2010}, respectively.
However, there is absence in literature of a systematic analytical approach applicable to the DCE in a relativistic system with a general oscillatory time-dependent parameter and that can be generalized to higher dimensions.
To fill this lack, in the present paper we develop a general perturbative analytical approach
which enables the investigation of additional bands in the spectrum of created particles via DCE, 
computing, systematically,  relativistic effects to any desired order. 
Although the main ideas of our approach are quite general, in the present work we shall focus on applications
related to the problem of the DCE in a superconducting circuit
\cite{J-R-Johansson-G-Johansson-C-Wilson-F-Nori-PRL-2009-2010,Wilson-Nature-2011,Silva-Farina-PRD-2011}.

The paper is organized as follows. 
In Sec. \ref{general-approach},
considering the SQUID experiment where the DCE is modeled by a Robin BC at a fixed point with a time-dependent Robin parameter,
we develop a general formula which gives the complete perturbative solution for the spectral distribution of the created particles.
In Sec. \ref{application-SQUID}, we use in the referred formula 
parameters related to the SQUID experiment, providing the experimentalists with some predictions.
Our final comments are in Sec. \ref{final-comments}.

\section{A general analytical approach for computing relativistic corrections}
\label{general-approach}

Let us start considering a superconducting coplanar waveguide 
with capacitance and inductance per unit length given, respectively, by $C_0$ and $L_0$,
and terminated through a SQUID (see Ref. \cite{J-R-Johansson-G-Johansson-C-Wilson-F-Nori-PRL-2009-2010}). 
Due to the presence of Josephson junctions in this system, according to \cite{J-R-Johansson-G-Johansson-C-Wilson-F-Nori-PRL-2009-2010} the electromagnetic field in this
coplanar waveguide can be conveniently described by the phase field operator $\phi(t,x)$,
defined by
$\phi(t,x)=\int^t dt^{\prime}E(t^{\prime},x)$,
where $E(t,x)$ is the electric field.
%
%
%
It can be shown (see details in Ref. \cite{J-R-Johansson-G-Johansson-C-Wilson-F-Nori-PRL-2009-2010}) that 
the phase field obeys the massless Klein-Gordon equation $(v^{-2}\;\partial^2_t-\partial^2_x)\phi\left(t,x\right) = 0,$
where $v=1/\sqrt{C_0L_0}$ is the velocity of light in
the wave\-guide (throughout this paper we shall assume $\hbar=v=1$). 
Applying appropriately Kirchoff's laws to the superconducting circuit, the authors in 
\cite{J-R-Johansson-G-Johansson-C-Wilson-F-Nori-PRL-2009-2010} show that the BC satisfied by $\phi$ is
the following
%
%
\begin{equation}
\phi\left(t,0\right)-\gamma\left(t\right) (\partial_x \phi)\left(t,0\right)\approx 0,
\label{time-dependent-Robin-BC}
\end{equation}
with
%
$
\gamma(t) = - \bar{\Phi}^2_0 \left[ {(2\pi)^2E_{J}(t)L_0} \right]^{-1}
=
- L_\text{eff}\left(t\right),
$
where $\bar\Phi_0$ is the magnetic quantum flux, $E_{J}\left(t\right)$ is the effective Josephson energy
(which depends on the magnetic flux), and $L_\text{eff}\left(t\right)$ is an effective length that modulates the change in time of the distance between the SQUID to an effective mirror at origin \cite{J-R-Johansson-G-Johansson-C-Wilson-F-Nori-PRL-2009-2010}.
The previous equation corresponds to a Robin BC at a fixed point with a time-dependent Robin parameter.
Considering the following general expression for the Josephson energy $E_{J}\left(t\right) =E_{J}^0 \left[1 + \epsilon f(t)\right]$,
where $0 < \epsilon < 1$ and $|f(t)|<1$, we can write
the time-dependent Robin parameter as
\begin{equation}
\gamma\left(t\right) \approx \gamma_0 \Bigl[ 1+ \sum_{k=1}^{N}{\epsilon^k f_k\!\left(t\right)}\Bigr],
\label{gamma-of-t}
\end{equation}
where $f_k\!\left(t\right)=\left[-f(t)\right]^k$,
$\gamma_0=- \bar{\Phi}^2_0 \left[ {(2\pi)^2 E_{J}^0 L_0} \right]^{-1}$
and $N$ denotes the order of expansion under investigation.

Now, we extend the Ford-Vilenkin perturbative approach \cite{Ford-Vilenkin-PRD-1982} by expressing
the field solution in the form
\begin{equation}
{\phi}\left( t,x\right) \approx
{\phi}_{0}\left( t,x\right) + \sum_{j=1}^{N}\epsilon ^{j}{\phi}_{j}\left( t,x\right),
\label{High-Orders-of-Ford-Vilenkin}
\end{equation}
where $\phi_{0}\left(t,x\right)$ is the unperturbed field and 
$\epsilon^j\phi_{j}\left(t,x\right)$ represents the correction of order $j$,
with $\phi_{0}$ and $\phi_j$ obeying the massless Klein-Gordon equation.
The unperturbed field $\phi_{0}$ obeys the usual static Robin
BC at origin,
$
\phi_{0}\left(t,0\right) = \gamma_0 \partial_x \phi_{0}\left(t,0\right).
$
As a consequence, the unperturbed field $\phi_{0}$ is given, for $x>0$,
$$
\phi_{0}\left(t,x\right)=\frac{1}{2\pi}
\int_{-\infty }^{\infty } \!\! dk \;\Phi_0(k,x) e^{-i\omega _{k}t},
$$
where
%
%
\begin{equation} 
\Phi_0(k,x) =
\sqrt{\frac{4\pi}{\omega _{k}\left(1+k^2\gamma_0^2\right)}}
g(k,x,\gamma_0)\left[a(k)\Theta \! \left( k\right) \!-\! {a}^{\dagger }\!(-k)\Theta \! \left( -k\right)\right]\!,
\label{static-Robin-BC-field-solution-transformada}
\end{equation}
with
\begin{equation}
g(k,x,\gamma_0)= \sin \left( k x\right) +k \gamma _{0}\cos \left( k x\right).
\label{eqforg}
\end{equation}
In the previous equations, 
$\Theta \! \left( k\right)$ is the Hea\-vi\-si\-de function, $a^{\dagger }(k)$  and $a(k)$ are the creation and annihilation operators that satisfy the canonical commutation rule
$[a(k),a^{\dagger}(k^{\prime})] = \delta \! \left( k - k^{\prime} \right)$ and $\omega_k=|k|$.
%
It can be shown that the fields $\phi_{j}$ obey the BC
\begin{equation}
\phi_{j}\left(t,0\right)-\gamma_0 \partial_{x}\phi_{j}\left(t,0\right) \approx
\gamma_0 \sum_{k=1}^j
{f_k\!\left(t\right)\partial_{x} \phi_{j-k}\left(t,0\right)}.
\label{High-order-Robin-BC}
\end{equation}
%
%
The time Fourier transform of the field $\phi_j\left(t,x\right)$, denoted by $\Phi_j(\omega,x)$, satisfies Helmholtz equation
$(\omega^2+\partial^2_x){\Phi}_{j}\left(\omega,x\right) = 0$.
Taking the Fourier transform of Eq. (\ref{High-order-Robin-BC}), we see that the field ${\Phi}_{j}\left( \omega ,0\right)$ satisfies to the following BC
\begin{equation}
\left[ 1 - \gamma_{0}\partial _{x} \right]{\Phi}_{j}\left( \omega ,0 \right) =
\!\!\int_{-\infty }^{\infty } \!\!\! d\xi\! \left[ \mathcal{O}_{\omega ,\xi }^{\left( j\right) } \! \left( x\right)
{\Phi}_{0} \! \left( \xi,x\right)\right]_{x=0},
\end{equation}
%
where
%
\begin{eqnarray}
\mathcal{O}_{\omega ,\xi }^{\left( j\right) }\!\left( x\right) &=&
\sum_{k=1}^j 
\int_{-\infty}^{\infty} \!\!\frac{d\xi_1}{2\pi} \frac{i\xi_1}{1-i\xi_1 \gamma_0}
F_{k}\!\left(\omega -\xi_1\right)\!\mathcal{O}_{\xi_1 ,\xi }^{\left( j-k\right) }\!\!\left( x\right),
\cr
\mathcal{O}_{\xi_1 ,\xi }^{\left( 0 \right) }\left( x\right) &=&
\left( i\xi_1 \right)^{-1} \left( 1-i\xi_1 \gamma_0 \right) \delta \! \left(\xi_1-\xi\right) \partial_{x},
\label{eq:recorrence}
\end{eqnarray}
%
%
%
represent recurrence formulas for the operators $\mathcal{O}_{\omega ,\xi }^{\left( j\right) }\left( x\right)$ that act on the spatial part of the unperturbed field ${\Phi}_0\left(\omega,x\right)$, and $F$ is the Fourier transform of $f$.
As causality requires, the solution ${\Phi}_{j}\left(\omega,x\right)$ of the Helmholtz equation must
lead to a solution $\phi_j\left(t,x\right)$ that travels from the mirror to infinity.
The time Fourier transform of $\phi\left(t,x\right)$ can be written as
%
\begin{equation}
{\Phi}\left(\omega,x\right) \approx
{\Phi}_{0}\left(\omega,x\right) + \sum_{j=1}^{N}\epsilon ^{j}{\Phi}_{j}\left(\omega,x\right).
\label{eq:fourier-formula-for-the-scalr-field}
\end{equation}
Considering that $\gamma(t\rightarrow-\infty)=\gamma_0$,
the in field $\Phi_0^{\text{in}}$ can be written as $\Phi_0$ in Eq.
(\ref{static-Robin-BC-field-solution-transformada}), with $a$ and $a^\dagger$ relabeled as $a_{\text{in}}$
and $a^{\dagger}_{\text{in}}$.
Ana\-lo\-gous\-ly, considering that $\gamma(t\rightarrow\infty)=\gamma_0$,
the out field $\Phi_0^{\text{out}}$ can be written as $\Phi_0$ in Eq.
(\ref{static-Robin-BC-field-solution-transformada}), with $a$ relabeled as $a^\text{out}$.
The in and out fields are then related by
\begin{equation}
{\Phi}_0^\text{out}\left( \omega ,x\right) =
{\Phi}_0^\text{in}\left( \omega ,x\right) - \frac{1}{\gamma_0}
\left[G^{\text{ret}}\left( \omega;x,0\right) - G_{\text{adv}}\left( \omega;x,0\right)\right]
\left[ 1 - \gamma_{0}\partial _{x} \right]
\Bigl[\sum_{j=1}^{N}\epsilon ^{j}{\Phi}_{j}\left( \omega,x\right)\Bigr]_{x=0},
\end{equation}
where the advanced  and retarded  Green functions are given by
%
\begin{equation}
G^\text{ret}_\text{adv} \left( \omega ;x,0 \right) =
\gamma_0 \left[1 \mp i\omega \gamma_0  \right]^{-1}e^{\pm i\omega x}.
\label{eq:retarded-and-advanced-Green-functions}
\end{equation}
Using Eq.(\ref{static-Robin-BC-field-solution-transformada}) and substituting the Green's function in
equation linking ${\Phi}_0^\text{out}$ and ${\Phi}_0^\text{in}$, we get the correspondent Bo\-go\-liu\-bov transformation
\begin{equation}
a_\text{out}(\omega ) = a_\text{in}(\omega ) +
\int_{-\infty }^{\infty }\!\!\! d\xi \sqrt{\frac{1}{|\xi| \left( 1+\xi ^{2}\gamma _{0}^{2}\right)}}
\sum_{j=1}^{N}\epsilon ^{j}
{\cal G}^{(j)}(\omega,\xi)
\left[ a_\text{in}(\xi) \Theta\!\left( \xi \right) -
a^{\dagger}_\text{in}(-\xi )\Theta\!\left( -\xi \right)\right],
\label{High-Order-Bogoliubov-Transformation}
\end{equation}
where
\begin{equation}
{\cal G}^{(j)}(\omega,\xi)=2i\sqrt{\frac{\omega}{1+\omega ^{2}\gamma_{0}^{2}}}\left\{\mathcal{O}_{\omega ,\xi }^{\left(j\right) }\left( x\right)
g(\xi,x,\gamma_0)\right\}_{x=0}.
\end{equation}
For the case $N=1$ we find
the correspondent formula found in Ref. \cite{Silva-Farina-PRD-2011}.
The linear dependence of $a_{\text{out}}$ in terms of $a^{\dagger }_{\text{in}}$ clearly indicates a non-vanishing particle creation distribution in the out state.

Since we are interested in computing the conversion
caused by $\gamma(t)$ of vacuum fluctuations into real field excitations,
we consider the vacuum ($\left.|0_{\text{in}}\right\rangle$) as the in state of the system.
From the Bogoliubov transformations (\ref{High-Order-Bogoliubov-Transformation}), we obtain a very general formula (in 1+1 dimensions) 
for the number of created particles between $\omega$ and $\omega+d\omega$ per unit frequency, namely,
%
%
\begin{widetext}
\begin{equation}
{\cal{N}}(\omega) =
\left\langle 0_{\text{in}}|\right. a^{\dag }_{\text{out}}(\omega)\; a_{\text{out}}(\omega )\left.|0_{\text{in}}\right\rangle
\approx
{\sum_{j,k=1}^{N\;\;\prime}}
\int_{-\infty}^{\infty } \!\! d\xi \;
\frac{\epsilon ^{j+k} \Theta \! \left( -\xi \right) \!}{|\xi| \left( 1+\xi ^{2}\gamma _{0}^{2}\right)}
{\cal G}^{(j)\ast}(\omega,\xi){\cal G}^{(k)}(\omega,\xi).
\label{High-order-spectral-distribution}
\end{equation}
\end{widetext}
where ${\sum_{j,k=1}^{N\;\;\prime}}$ means that the indices $j$
and $k$ are chosen so that $j+k\leq N+1$.
In other words, Eq. (\ref{High-order-spectral-distribution}) gives
the perturbative solution for the spectral distribution $\cal{N}(\omega)$
up to order $N+1$ in $\epsilon$, and for an arbitrary time-dependence of the Robin parameter
given according to Eq. (\ref{gamma-of-t}).

\section{Application to the SQUID experiment}
\label{application-SQUID}

\begin{figure}[!hbt]
\centering
\includegraphics[scale=0.38]{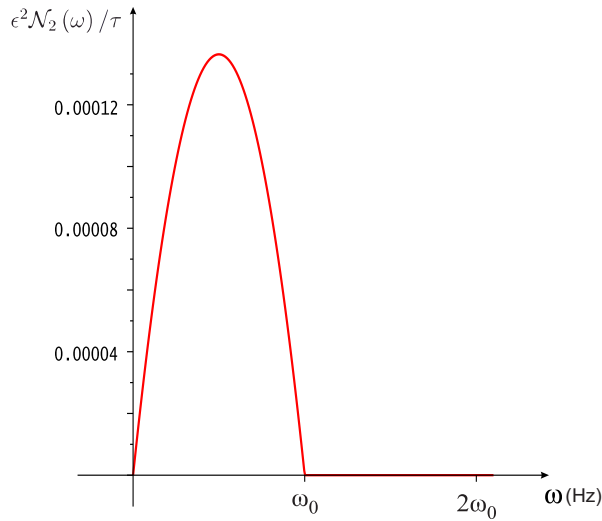}
\caption{(color online)
First (non-relativistic) parabolic band $\epsilon^2\mathcal{N}_{2}(\omega)/\tau$.
%
}
\label{epsilon-2-N-2}
\end{figure}

Hereafter we consider, for practical purposes, a particular application of the formula
(\ref{High-order-spectral-distribution}) for a typical oscillatory time variation of the Robin parameter given by
$
f\!\left(t\right) = \cos \left( \omega_0 t \right) e^{-|{t}|/ \tau},
$
with $\omega_0\tau\gg1$ (monochromatic limit) \cite{J-R-Johansson-G-Johansson-C-Wilson-F-Nori-PRL-2009-2010,Silva-Farina-PRD-2011}, where $\omega_0$ is the characteristic frequency and $\tau$ is the effective time interval in which the oscillations occur.
In this context, Eq. (\ref{High-order-spectral-distribution})
requires the solution of integrals
having the general form
$\int_{-\infty}^{\infty} d\omega A(\omega) B(\omega,\omega_0\tau,\omega_i)$,
where, as $\omega_0\tau\rightarrow\infty$, the functions $B(\omega,\omega_0\tau,\omega_i{\text{'s}})$
exhibit sharped peaks at $\omega=\omega_i{\text{'s}}$, so that, in solving these integrals, we can apply
\begin{equation}
\lim_{\omega_0\tau\rightarrow\infty}
\int_{-\infty}^{\infty} d\omega A(\omega) B(\omega,\omega_0\tau,\omega_i) =
\left[\sum_iA(\omega_i)\right]
\int_{-\infty}^{\infty} d\omega B(\omega,\omega_0\tau,\omega_i{\text{'s}}).
\label{eq:delta-int}
\end{equation}
%
%
Hereafter we also consider the following parameters related to the SQUID experiment:
$\omega_0=2\pi \times 10.30$ GHz, $\epsilon=0.25 $, $v = 1.2 \times 10^{8}$ m/s and $\gamma_0 = 0.44 \times 10^{-3}$ m \cite{J-R-Johansson-G-Johansson-C-Wilson-F-Nori-PRL-2009-2010, Wilson-Nature-2011}.
For $N=1$, Eq. (\ref{High-order-spectral-distribution}) gives the non-relativistic analytical result
${\cal{N}}(\omega) \approx \epsilon ^{2}\mathcal{N}_{2}\! \left( \omega \right)$,
with the analytical formula for $\mathcal{N}_{2}\! \left( \omega \right)$ being the same as that
found in Ref. \cite{Silva-Farina-PRD-2011}, and whose parabolic behavior is showed in Fig. \ref{epsilon-2-N-2},
where we can observe that the spectrum is null for $\omega>\omega_0$. 
%
For $N=2$, we get a negligible correction (proportional to $\epsilon^3$) to the parabolic first band in the spectrum
($\mathcal{N}_{3}\approx 0$).

\begin{figure}[!hbt]
\centering
\includegraphics[scale=0.38]{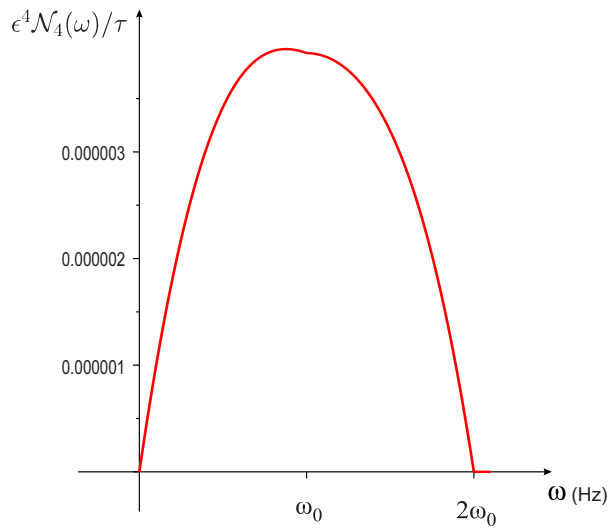}
\caption{(color online)
First relativistic correction $\epsilon^4\mathcal{N}_{4}(\omega)/\tau$.
%
}
\label{epsilon-4-N-4}
\end{figure}

\begin{figure}[!hbt]
\centering
\includegraphics[scale=0.38]{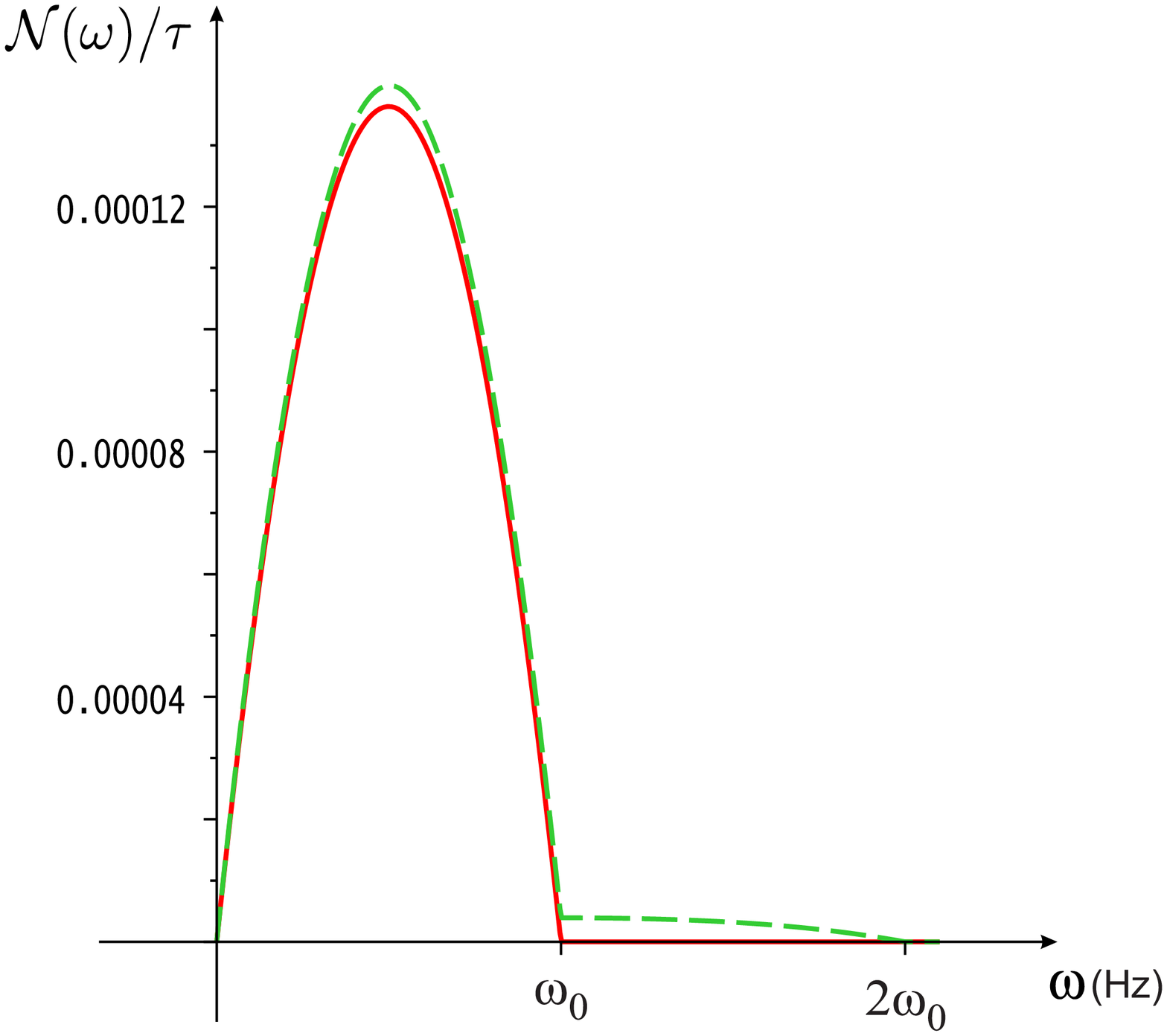}
\caption{(color online)
The solid line is the first (non-relativistic) parabolic band $\epsilon^2\mathcal{N}_{2}(\omega)/\tau$.
The dashed line is $\mathcal{N}(\omega)=\left[\epsilon^2\mathcal{N}_2(\omega)+\epsilon^4\mathcal{N}_{4}(\omega)\right]/\tau$.
%
}
\label{epsilon-2-N-2-epsilon-4-N-4}
\end{figure}

Now, let us include the relativistic corrections. Considering $N=3$, we get several integrals to be solved and,
after several cumbersome calculations, we get a long final analytical formula for
${\cal{N}}(\omega)$, which can be represented by
$
{\cal{N}}(\omega) \approx \epsilon ^{2}\mathcal{N}_{2}\! \left( \omega \right)
+\epsilon ^{4}\mathcal{N}_{4}\! \left( \omega \right),
$
where 
$\epsilon^4\mathcal{N}_{4}\! \left( \omega \right)$ describes the relativistic corrections in the spectral density.
Since the analytical formula for $\mathcal{N}_{4}(\omega)$ is too long to be written here, 
we just exhibit its graphical behavior in Fig. \ref{epsilon-4-N-4} (solid line).
We observe that $\mathcal{N}_{4}(\omega)$ is null for $\omega>2\omega_0$.
Notice that $\mathcal{N}_{4}$ contributes either to correct the non-relativistic band
$\epsilon^2\mathcal{N}_{2}$ in the region $0<\omega<\omega_0$, as to compose the additional
part in the region $\omega_0<\omega<2\omega_0$ of the spectrum.
A subtle difference between our result and the one found in Ref. \cite{J-R-Johansson-G-Johansson-C-Wilson-F-Nori-PRL-2009-2010} is that the additional band $\epsilon^4\mathcal{N}_{4}$ does not have exactly the parabolic form going from 0 to $2\omega_0$, as considered in \cite{J-R-Johansson-G-Johansson-C-Wilson-F-Nori-PRL-2009-2010}.
After adding $\epsilon^2\mathcal{N}_{2}$ with $\epsilon^4\mathcal{N}_{4}$, 
we get the shape shown in Fig. \ref{epsilon-2-N-2-epsilon-4-N-4} (dashed line),
which is in good agreement with the one obtained via numerical methods in Ref. \cite{J-R-Johansson-G-Johansson-C-Wilson-F-Nori-PRL-2009-2010}.

Let us now provide the experimentalists with some predictions.
From our calculations, we obtain that, in the non-relativistic approximation ($\epsilon^2\mathcal{N}_{2}$), 
the rate of photon creation is $5.8\times10^6/\text{sec}$,
and the maximum frequency is $\omega=\omega_0$. Taking into
account the first relativistic correction ($\epsilon^2\mathcal{N}_{2}+\epsilon^4\mathcal{N}_{4}$),
the range of frequencies of the created particles is extended up to $2\omega_0$, 
the rate of photon creation is enhanced to $6.2\times10^6/\text{sec}$ ($6\%$ greater in relation to the
non-relativistic result), and the rate of created particles with frequencies $\omega_0<\omega<2\omega_0$ 
(relativistic band) is $1.8\times10^5/\text{sec}$. These photons with frequencies in the interval $\omega_0<\omega<2\omega_0$
may, in principle, be observed, providing this way an extra signatures for the identification of the dynamical Casimir effect.

\begin{figure}[!hbt]
\centering
\includegraphics[scale=0.38]{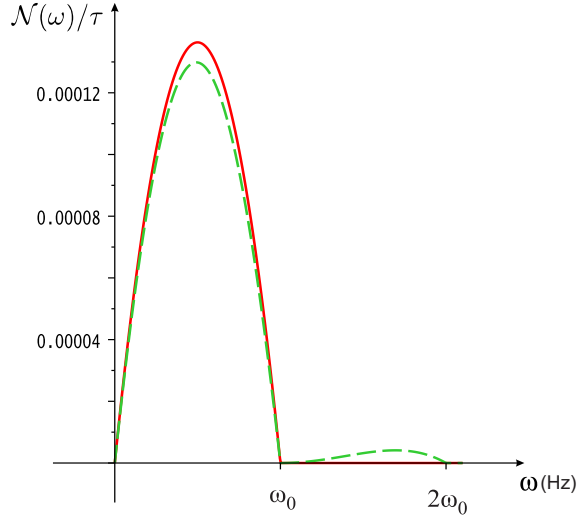}
\caption{(color online)
The solid line is the first (non-relativistic) parabolic band $\epsilon^2\mathcal{N}_{2}(\omega)/\tau$.
The dashed line is $\mathcal{N}(\omega)=\left[\epsilon^2\mathcal{N}_2(\omega)+\epsilon^4\mathcal{N}_{4}(\omega)\right]/\tau$.
%
}
\label{epsilon-2-N-2-epsilon-4-N-(1)-4}
\end{figure}

Finally, as an additional check for our analytical formulas, 
we investigate the toy model in which we prescribe the time behavior 
of the parameter $\gamma(t)$ such that it is exactly
given by $\gamma\left(t\right) =\gamma_0 [ 1+ {\epsilon f_1\!\left(t\right)}]$.
It is possible to show that this model approximately describes a moving mirror imposing a Dirichlet boundary condition 
to the field. Now, the function $\mathcal{N}_{4}(\omega)$ is relabeled as $\mathcal{N}_{4}^{(1)}(\omega)$,
and, after adding $\epsilon^2\mathcal{N}_{2}$ with $\epsilon^4\mathcal{N}_{4}^{(1)}$, 
we get the shape shown in Fig. \ref{epsilon-2-N-2-epsilon-4-N-(1)-4} (dashed line).
This shape is in agreement with the prediction (see Ref. \cite{Lambrecht-et-al-EPJD-1998})
of additional bands vanishing for all frequencies $\omega$
equal to an integer multiple of $\omega_0$, with the spectrum decomposed into a succession
of arches, each one limited by two successive multiples of $\omega_0$, and with
two points of maximum: one of them slightly shifted to the left
in relation to $\omega_0/2$, and the other one shifted to the right
in relation to $3\omega_0/2$. 
%
%
%

\section{Final remarks}
\label{final-comments}

In this work we presented an analytical approach to the DCE which includes systematically
relativistic effects to any desired order. We applied our method to the model which
describes theoretically the setup used  in the SQUID experiment \cite{J-R-Johansson-G-Johansson-C-Wilson-F-Nori-PRL-2009-2010}.
We made estimatives of the first relativistic effects, providing the experimentalists with
theoretical predictions to be tested. Particularly, we estimated the intensity of the first relativistic band
(relativistic effect) relative to the first one (non-relativistic result).
It is very important to take into account these corrections because they may provide an extra signature for
identifying the dynamical Casimir photons. The observation of an additional frequency band would be
a remarkable experimental achievement.

Finally, we remark that the generalization of the Ford-Vilenkin approach \cite{Ford-Vilenkin-PRD-1982},
shown in the present work, to investigate the appearance of additional
bands in the spectral density, considering the theoretical model underlying the SQUID experiment,
requires not only the generalization given by the Eq. (\ref{High-Orders-of-Ford-Vilenkin}),
but also includes (\ref{gamma-of-t}). Only taking into account both generalizations, one can
reproduce the bands shown in Figs. \ref{epsilon-4-N-4} (solid line) and \ref{epsilon-2-N-2-epsilon-4-N-4}
(dashed line), in agreement with the numerical results found in Ref. \cite{J-R-Johansson-G-Johansson-C-Wilson-F-Nori-PRL-2009-2010}.
Though our calculations were made in 1+1 dimensions, because we were interested in discussing the
SQUID experiment, our approach can be generalized to other systems, higher dimensions,
dif\-fe\-rent fields (electromagnetic fields, massive fields...) and boundary conditions.

\section{Acknowledgments}
\label{Ack}
We thank J.D.L. Silva, A.N. Braga, P.A. Maia Neto and C.M. Wilson for valuable discussions. This work was partially supported by CNPq, CAPES and FAPERJ.



\begin{thebibliography}{99}

\bibitem{Moore-1970} G.T. Moore, {J Math. Phys.} {\bf 11}, 2679 (1970).
%
\bibitem{Dewitt-PhysRep-1975} B.S. DeWitt, {Phys. Rep.} {\bf 19}, 295 (1975).
%
\bibitem{Fulling-Davies-PRS-1976-I} S.A. Fulling and P.C.W. Davies, {Proc. R. Soc. London} {\bf A 348}, 393 (1976).
%
\bibitem{Fulling-Davies-PRS-1977-I} P.C.W. Davies and S.A. Fulling, {Proc. R. Soc. London} {\bf A 354}, 59 (1977).
%
\bibitem{Lambrecht-et-al-EPJD-1998} A. Lambrecht {\it et al}, {Eur. Phys. J. D}  {\bf 3}, 95 (1998).
%
\bibitem{Dodonov-PRA-1996} V.V. Dodonov and A.B. Klimov, Phys. Rev. A,
{\bf 53}, 2664 (1996).
%
\bibitem{Yablonovitch-1989}  E. Yablonovitch, {Phys. Rev. Lett.} {\bf 62}, 1742 (1989);
%
\bibitem{Braggio-Agnesi}C. Braggio {\it et al}, {Europhys. Lett} {\bf 70}, 754 (2005);
A. Agnesi {\it et al}, {J. Phys.} A {\bf 41}, 164024 (2008);
A. Agnesi {\it et al}, {J. Phys: Conf. Series} {\bf 161}, 012028 (2009).
%
\bibitem{J-R-Johansson-G-Johansson-C-Wilson-F-Nori-PRL-2009-2010} J.R. Jo\-han\-sson, G. Jo\-han\-sson, C.M. Wilson and F. Nori, {Phys. Rev. Lett.} {\bf 103}, 147003 (2009); J.R. Jo\-han\-sson, G. Jo\-han\-sson, C.M. Wilson and F. Nori, {Phys. Rev. A}  {\bf 82}, 052509 (2010).
%
%
\bibitem{Dezael-Lambrecht-EPL-2010} F.X. Dezael and A. Lambrecht, {Eur. Phys. Lett.} {\bf 89}, 14001 (2010).
%
\bibitem{Kawakubo-Yamamoto-PRA-2011} T. Kawakubo and K. Yamamoto, {Phys. Rev.} A, {\bf 83}, 013819 (2011).
%
\bibitem{Faccio-Carusotto-EPL-2011}  D. Faccio and I. Carusotto, {Eur. Phys. Lett} {\bf 96}, 24006 (2011).
%
\bibitem{Kim-Brownell-Onofrio-2006} W.J. Kim, J.H. Brownell and R. Onofrio , {Phys. Rev. Lett.} {\bf 96}, 200402 (2006).
%
\bibitem{Wilson-Nature-2011} C.M. Wilson {\it et al}, {Nature (London)} {\bf 479}, 376 (2011).
%
\bibitem{Lambrecht-Jaekel-Reynaud-PRL-1996} A. Lambrecht, M.T. Jaekel and S. Reynaud, {Phys. Rev. Lett.}  {\bf 77}, 615 (1996).
%
\bibitem{Mintz-Farina-Maia-Neto-Robson-JPA-2006} B. Mintz {\it et al}, {J. Phys.} A {\bf 39}, 6559 (2006); B. Mintz {\it et al}, { J. Phys.} A {\bf 39}, 11325 (2006).
%
\bibitem{Rego-Mintz-Farina-Alves-PRD-2013} Andreson L. C. Rego, B.W. Mintz, C. Farina and D.T. Alves, { Phys. Rev.} D {\bf 87}, 045024 (2013).
%
\bibitem{Silva-Farina-PRD-2011} H. O. Silva and C. Farina, {Phys. Rev. D} {\bf 84}, 045003 (2011).
%
\bibitem{Fosco-et-al-PRD-2013} C.D. Fosco, F.C. Lombardo and F.D. Mazzitelli,  {Phys. Rev. D} {\bf 87}, 105008 (2013).
%
\bibitem{Ford-Vilenkin-PRD-1982} L.H. Ford and A. Vilenkin, {Phys. Rev. D} {\bf 25}, 2569 (1982).
%
\end{thebibliography}
\end{document}